\documentclass[12pt,twoside,draftclsnofoot,onecolumn]{IEEEtran}

\ifCLASSINFOpdf
\else
\fi

\usepackage{graphicx,subfigure,array,amssymb,latexsym,amssymb,lipsum,algpseudocode,epsfig,float,epsf,amsmath,epstopdf,cite,tikz, float,algpascal,amsthm, bm, stackengine,multicol,microtype}
\usepackage{dblfloatfix}
\usepackage[linesnumbered,ruled,vlined]{algorithm2e}
\usepackage{cuted}
\DeclareGraphicsExtensions{.pdf,.png,.jpg,.eps}

\makeatletter
\newcommand{\removelatexerror}{\let\@latex@error\@gobble}
\makeatother

\newtheorem{rem}{Remark}

\newtheorem{defin}{Definition}
\SetKwProg{Fn}{Function}{}{}
\SetKwRepeat{Do}{do}{while}
\newcommand\norm[1]{\left\lVert#1\right\rVert}
\usepackage{booktabs}
\usepackage{multirow}

\graphicspath {{/Simulation/Figure/}}

\begin{document}
%




\title{User-Access Point Association for High Density MIMO Wireless LANs}

\author{{\IEEEauthorblockN{Phillip B. Oni and Steven D. Blostein \\ Dept. of Electrical \& Computer Engineering\\ Queen's University, Kingston, ON.}}}
\maketitle

\begin{abstract}

\noindent Wireless local area network (WLAN) access points (APs) are being deployed in high density to improve coverage and throughput. The emerging multiple-input multiple-output (MIMO) implementation for uplink (UL) transmissions promises high per-user throughput and improved aggregate network throughput. However, the high throughput potential of dense UL-MIMO WLAN is impaired by multiple access channel interference and high contention among densely distributed user stations (STAs). We investigate the problem of actualizing  the throughput potential of UL-MIMO in high density WLANs via user-AP association. Since user-AP association influences interference and STA contention, a method to optimally distribute STAs among APs is proposed to maximize aggregate users' throughput utility. This problem is transformed into a graph matching problem with the throughput utility function as the graph edge weights. The graph matching problem is solved as a combinatorial problem using a modified classical Kuhn-Munkres algorithm. A dynamic implementation of the proposed algorithm is used to periodically update user-AP associations when there are changes in the network due to new entrants and/or user mobility. Simulated dense UL-MIMO WLAN scenarios reveal that the proposed scheme achieves an average of  $36.9 \%$, $33.5 \%$, $20.4 \%$ and $11.3 \%$  gains over the default strongest signal first (SSF) association scheme used in conventional WLAN, \textit{Greedy} \cite{yzhangd}, \textit{SmartAssoc} \cite{smartasoc} and \textit{best performance first} (BPF) \cite{weili} algorithms, respectively.





\end{abstract}



\section{Introduction}

The availability of high data rates in a wireless local area network (WLAN) at low cost and wide spread WiFi-enabled devices result in dense deployments of access points (APs) in multi-tenant residential areas, enterprise buildings, indoor/outdoor hotspots in hotels, airports, and caf\'es.  Dense AP deployment is also envisaged to facilitate cellular-WiFi data offloading. Consequently, as more APs are closely spaced, contention and interference \cite{shin} among user stations (STAs) and APs inevitably increase. While a dense wireless local area network (DWLAN) is able to guarantee coverage, performance degradation becomes severe due to increased interference and STA contention, particularly in view of the limited number of available orthogonal channels. Therefore, the main challenges for WLAN systems include interference management and resource sharing. Enhancing user throughput in the presence of severe interference is challenging mostly due to the randomness of both users activities (or traffics) and wireless channel. 

The inherent high contention and interference in large-scale WLAN necessitate the need for new techniques to improve performance in the face of growing user density. To enhance the uplink throughput in a dense environment, emerging WLAN standard, 802.11ax \cite{ieeeax1}, \cite{ieeeax2}, \cite{ieeeax3} supports uplink multiuser (MU)-MIMO. However, as the density of user stations increases, interference becomes severe and user throughput could degrade significantly. This is mainly because dense networks have multiple overlapping basic service sets (BSSs) (analogous to cells in cellular networks), which causes inter-cell or inter-BSS interference and contention. Arguably, interference distribution depend on user-AP association and optimally distributing users among APs considering the interference level experienced by each user at the associated AP, could improve performance in an interference-dominated WLAN. For user-AP association, current IEEE 802.11 WLAN standards employ a {\em strongest signal first} (SSF) approach, which associates an STA with the AP based on the strongest received signal strength (RSS) without considering interference levels more globally, recognizing that severe interference and contention at some APs may degrade aggregate network throughput \cite{weili}. In other words, associating with the SSF AP (possibly the one that is physically closest) does not guarantee best overall performance \cite{weili}, \cite{ref4}. 

The inefficiency of the conventional SSF scheme motivated several studies \cite{weili} - \cite{imad} in search for a user-AP association technique that improves different performance metrics in wireless LANs, including proportional fairness. A special case of achieving proportional fairness via AP association appears in \cite{weili}, where the authors study proportional fairness in multi-rate WLANs using AP association. The \textit{best performance first} (BPF) algorithm \cite{weili} achieves proportionally fair performance for single-input single-output (SISO) case without the consideration of the effects of MAC protocols on throughput. Similarly, in \cite{pbo}, we proposed graph algorithm to improve uplink (UL) throughput assuming SISO channel and without considering the impacts of MAC protocols. Our previous work in \cite{psdb} focuses on decentralizing user-AP association decision with the objective of improving downlink (DL) throughput in the presence of multi-AP interference. Therein, using pilot signals over a SISO channel, an STA measures inter-BSS interference at the target APs and select an AP that offers the best DL signal-to-interference plus noise ratio (SINR). The paradigm of wireless LAN is shifting towards support for multiple streams with multiple antennas at the users \cite{ieeeax3}. User-AP association needs to be configured to consider channel correlation between STAs and the target APs. 

Considering the effect of interference or SINR on AP association decisions and/or WLAN performance is gaining traction. An online algorithm that considers interference when selecting an AP appears in \cite{smartasoc}. In \cite{ref20}, user-AP association optimization minimizes the effects of inter-network interference in WLAN deployments. Karimi et al. \cite{ref26} seek to obtain proportional-fair AP association in dense networks with an assumption that a set of networks share the same upstream provider. Such cooperation among APs assumes that the APs belong to specific service provider, which is not the case in scenarios like multi-tenant networks \cite{wzhang}. Yen et. al \cite{yen} study a \textit{game theoretic} STA-AP association control that jointly considers stability and fairness. Performance based mainly on the PHY rate measured in terms of the SINR does not account for the MAC protocol effects such as delays due to collision, retransmissions and contention, which are envisaged to be severe in ultra dense WLANs. New association schemes that jointly consider interference level and MAC protocol effects are desirable for large-scale multi-antenna WLANs.

In \cite{yzhangd}, the authors propose two association algorithms and provide thorough analysis of the performance bounds. The  \textit{greedy algorithm} \cite{yzhangd} allows a user to selfishly associate with an AP with low load upon entering the network and stay associated with the same AP for the entire time spent in the network. Using the same least loaded AP decision metric, users select AP with least load in \textit{the best response algorithm} but users are allowed to switch their AP multiple times. Other load-based schemes are proposed in \cite{zhayang}, \cite{leiyou} and \cite{qyerong}. These algorithms only consider the load at each AP while neglecting the MAC protocol effects on end-to-end throughput and the interference level at the target AP, whereas severe interference could significantly impact throughput and cause user payloads to collide, causing increased packet error rate and retransmission requests. 

For heterogeneous networks, the User-AP Association problem is addressed  in \cite{qyerong} with the primary objective of load balancing. Therein, the problem is formulated as a general utility maximization problem. User-AP association is optimized for load balancing and fairness in millimeter-wave (mmWave) wireless networks in \cite{gAthan}. The problem of user-AP association is formulated as an optimization problem with constraints on the demanded data rate of each client to minimize the maximum AP utilization. The authors of \cite{yuzheXu} addressed the joint association control and relaying  problem. The authors design distributed auction algorithms that find optimal client-relay-AP association. In \cite{ahmed}, the authors' ultimate objective is to optimize the user association to maximize the mean rate utility of full-duplex (FD) networks. \cite{ahmed} focuses primarily on cellular networks where base stations (BSs) are distributed according to Poisson Point Processes (PPPs) in the in-band FD mode. Similarly, \cite{yinjunL} addresses the joint user association and spectrum allocation problem for heterogeneous networks where the pico BSs operate in the in-band FD mode. User-AP association problem is approached differently in \cite{bethanab} for massive MIMO wireless networks. The authors formulate the problem as a convex network utility maximization with the assumption that the user rates are deterministic under certain channel conditions.

The effects of CSMA/CA or MAC protocol in high density WLANs will be severe given the high density of nodes contending for the limited radio resources. The above existing works do not consider this factor in the formulation of the actual system throughput. In other words, the delays induced by MAC protocol needs to be taken into account when making AP selection decision; this will guarantee that users are matched to APs for better end-to-end performance and quality of service (QoS). Moreover, DWLAN users are typically unevenly distributed among APs causing high contention and intra-cell interference at some APs regardless of the AP loads. For UL-MIMO nodes, using AP load as the decision metric could result in assignment of users to APs with bad channel correlation. Distributing users among APs based on some level of user-AP channel correlation and the achievable SINR of each user is feasible given the notion that each user has the capability to measure the SINR at the target APs using pilot signals \cite{qyerong} or preambles to obtain the channel side information (CSI). 

In this contribution, the user-AP association problem is formulated to consider the CSI, the SINR at the target APs, the MAC protocol effects and proportional fairness. Primarily, the user-AP association optimization problem is formulated as a sum logarithmic throughput utility maximization problem. Then we show that this problem can be transformed into a maximum weighted bipartite graph matching problem where the throughput utility function represents the graph  edge weights. Since, user-AP association problem is combinatorial, the solution to this graph matching problem is realized  through application of the classical Kuhn-Munkres algorithm (KMA). Transforming the problem to a graph-matching problem makes it possible to solve the user-AP association problem combinatorially. We study the performance of the proposed scheme in uplink (UL) network scenario because interference and contention are more severe in the UL of high density network with large number of contending non-cooperating STAs. The proposed technique can be implemented or generalized to maximize a downlink (DL) throughput objective function such as the DL throughput objective we formulated previously in \cite{psdb}. Assuming sufficient network coordination, the proposed solution can be implemented in a distributed manner, provided STAs have network-wide (or global) CSI to estimate interference from other concurrent transmitters; perfect CSI is often available through \textit{channel sounding} or transmission of \textit{pilot signals} or \textit{preambles}.  The main contributions in this paper are highlighted as follows:

\begin{itemize}
\item We formulate the aggregate proportional fair throughput utility to capture the CSI, the effect of interference at the target AP and the MAC layer protocol. This throughput utility objective is formulated as a logarithmic utility function that takes into account, the physical phenomenon of the channel (CSI) between STAs and APs, the fairness and the effect of MAC protocol. Such objective function considers that user throughput is not only affected by the channel characteristics measured in terms of SINR, but also the proportional fairness of the entire system and the inevitable MAC protocol that governs channel access in WLANs.

\item Due to the combinatorial nature of user-AP association problem,  we developed a \textit{combinatorial search} algorithm based on the classical KMA to search for the optimal user-AP association set that maximizes the throughput utility function. In other words, we treat the aggregate throughput utility maximization problem as a graph-matching problem whose optimal solutions exist through the use of the proposed \textit{combinatorial search} algorithm. To handle dynamic network scenarios, we design a dynamic implementation of the proposed scheme to update user-AP association set when network changes due to new entrants and/or user mobility.
		
\item Motivated by the timely development of a standard (802.11ax) supporting UL MIMO for WLANs, the performance of the proposed algorithm is examined using UL throughput as a metric where STAs are equipped with multiple antennas. For performance benchmarking in a WLAN with UL MIMO capability, network-simulated scenarios compare the performance of the proposed framework with that of the default SSF scheme used in conventional WLAN systems, the best performance first (BPF) \cite{weili}, the \textit{greedy algorithm} \cite{yzhangd} and the \textit{SmartAssoc algorithm} \cite{smartasoc}. The proposed framework achieves better aggregate throughput performance while increasing the density (in terms of users size) of the network.



\end{itemize}

The remaining sections in this paper are organized as follows. Section~\ref{systems} presents the UL-MIMO WLAN system model while the proposed framework is presented in Section~\ref{framework}. The performance evaluation is discussed in Section~\ref{performanceI} and Section~\ref{conclusion} provides the conclusions.

\section{System and Network Models}\label{systems}

In this section, we discuss the system assumptions and models used in our proposed user-AP association framework.

\subsection{Network Topology Model}

\begin{figure}[!h]
	\centering
	\includegraphics[width=3in]{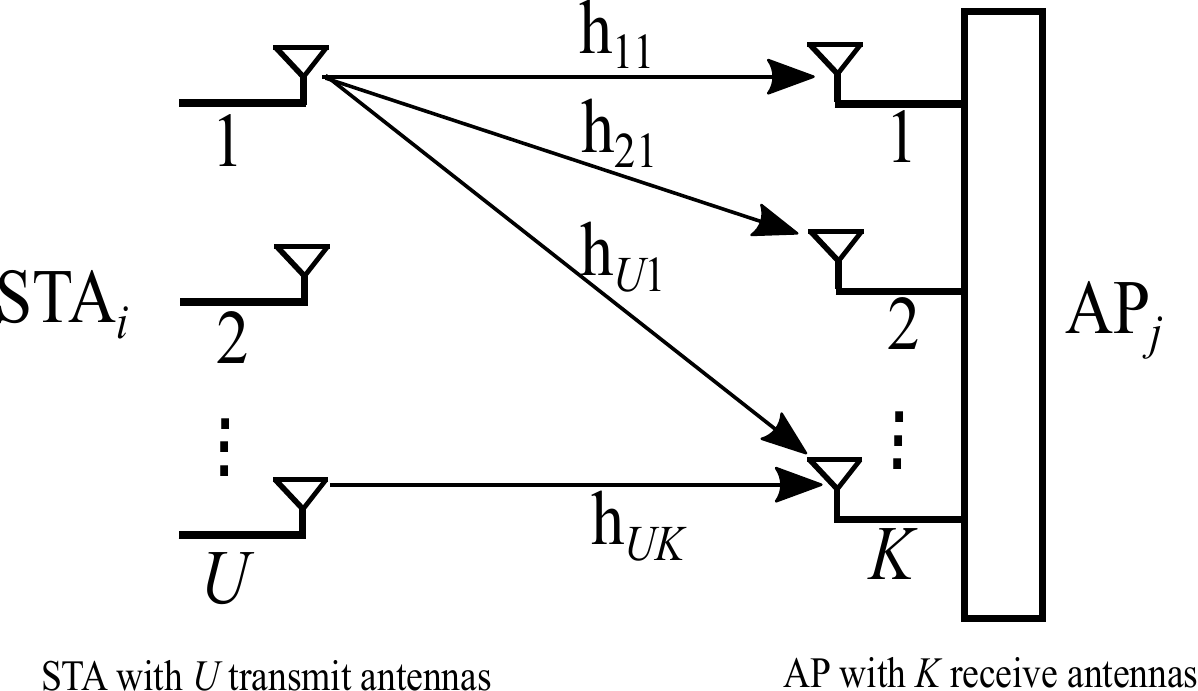}
	\caption{Uplink MU-MIMO model: $U$ antennas STA and $K$ antennas AP.}
	\label{fig:figMU}
\end{figure}

Let ${\mathcal{A}}$ represent the set of APs and ${\mathcal{N}}$ the set of STAs in the network. Let index STA$_i$ $\in {\mathcal{N}} $ as $1 \leq i \leq |{\mathcal{N}}| \equiv N $  and AP$_j$ $ \in {\mathcal{A}} $ as $1 \leq j \leq |{\mathcal{A}}|\equiv M$. The dense WLAN considered in this work is a multi-cell WLAN where each AP and its associated STAs form a basic service set (BSS) or cell, assuming that each STA associates with the closest AP based on the strongest received signal strength. As shown in Figure~\ref{fig:figMU}, each STA is equipped with $U$ antennas while each AP has $K$ receiving antennas.

\subsection{Uplink MIMO Channel Model}

Figure~\ref{fig:figMU} depicts MIMO channel model in a given BSS or cell; this system model is motivated by the emerging support for simultaneous uplink transmissions in 802.11ax \cite{ieeeax1},\cite{ieeeax2}, \cite{ieeeax3}. In Figure~\ref{fig:figMU}, let h$_{uk}$ denote the channel gain between $u$th transmit antenna element of STA$_i$ and the $k$th receive antenna element of AP$_j$, h$_{uk}$ is the $\left(u, k\right)\mbox{th}$ entry of the $U \times K$ channel matrix $\mathbf{H}_{ij}$; i.e., the narrowband channel between the $u^{th}$ transmit antenna at STA$_i$ and the $k^{th}$ receive antenna at AP$_j$. Let $\mathbf{x}$ represent the uplink transmitted signal vector from STA$_i$ to AP$_j$, and $\mathbf{y}$ is the received signal vector at AP$_j$. Assuming a quasi-stationary MIMO channel $\mathbf{H}_{ij}$ between STA$_i$ and AP$_j$ is known at the transmitter, the received signal vector at AP$_j$ can be expressed as
\begin{equation}
\mathbf{y} = \sqrt{\frac{ E_x }{ U}} \mathbf{W}_{ij} \mathbf{H}_{ij} \mathbf{x} + \mathbf{W}_{ij}\mathbf{n}, \qquad \forall i \in \mathcal{N},  \forall j \in \mathcal{A},
\end{equation}

\noindent where $E_x$ is the transmitted symbol energy, $\mathbf{W}_{ij}$ is the beamformer weight matrix whose vector entries $\mathbf{w} = \frac{\mathbf{h}^H}{\norm{ \mathbf{h} }}$ where $\mathbf{h} =  \left[h_{1}, h_{2} \cdots h_{u}\right] $ (MISO channel vector),  $\mathbf{n}$ represents the noise vector at AP$_j$ whose entries are $n_1 \cdots n_K$ zero mean white Gaussian noise with variance $\sigma_n^2$ at the $K$ receive antennas. The detection of the desired signal $\hat{\mathbf{x}}$ is achievable through channel inversion using the beamformer matrix $\mathbf{W}_{ij}$, which is obtainable from the channel knowledge $\mathbf{H}_{ij}$ as
\begin{equation}
\mathbf{W}_{ij} = \left( \mathbf{H}_{ij}^H \mathbf{H}_{ij} \right)^{-1} \mathbf{H}_{ij}^H, 
\label{beamer}
\end{equation}
\noindent where $\left( \cdot \right)^H$ denotes the Hermitian transpose of $\mathbf{H}_{ij}$. 


Based on the operation of the CSMA/CA protocol, only a subset of ${\mathcal{N}} $ STAs will gain access to the channel at a given time-slot, let ${\mathcal{N}}_{\mbox{csma}} $ denotes this subset of STAs permitted by the CSMA/CA protocol to transmit concurrently on the network. The signal-to-interference plus noise, $\mbox{SINR}_{ij}$ at AP$_j$ on interference-plus-noise-dominated channel bandwidth $B$, depends on the total interference power received by AP$_j$ from other concurrent transmitting STAs other than desired STA$_i$ is:
\begin{equation}
\mbox{SINR}_{ij} = \frac{\sqrt{\frac{ E_x }{ U}} |\mathbf{W}_{ij} \mathbf{H}_{ij}|^2  }{ \norm{\mathbf{W}_{ij}}^2 \sigma_n^2 + \sum\limits_{z \in {\mathcal{N}}_{\mbox{csma}} , z \neq i}^{} |\mathbf{W}_z \mathbf{H}_z|^2},
\label{sinrmeasure}
\end{equation}
\noindent where $\mathbf{W}_{ij}$ is given in (\ref{beamer}) and $\mathbf{H}_z$ is the channel between AP receiver and the interference transmitter $z$ with a corresponding beamformer $\mathbf{W}_{z}$ , the bit rate of the channel is given as
\begin{equation}
r_{ij} = B\cdot \log_2 \left( 1 + \mbox{SINR}_{ij}\right),
\label{ratePHY}
\end{equation}
\noindent where the receiving AP can estimate the total interference power received from STAs transmitting in  other BSSs. The set ${\mathcal{N}}_{\mbox{csma}}$ depends on the carrier sensing range (CSR) \cite{cthorpe} that determines the degree of spatial reuse.

\begin{figure*}[!b]
	\centering
	\includegraphics[width=5in ]{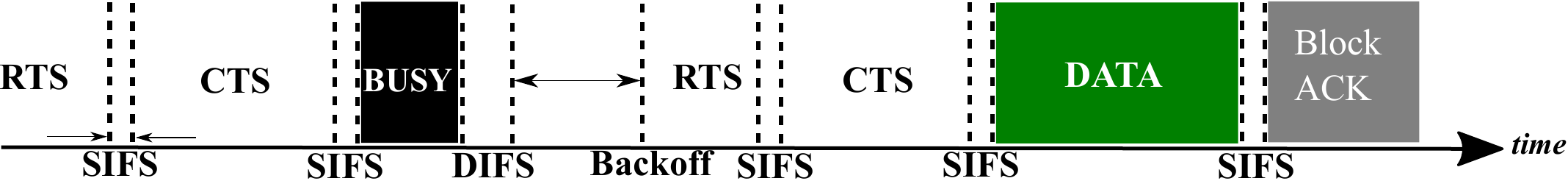}
	\caption{Channel negotiation procedure.}
	\label{fig:figTimer}
\end{figure*}

\subsection{Effective User Throughput} {\label{effectiveThru}}

Given the channel rate $r_{ij}$  of STA$_i$'s  on the network when associated with AP$_j$. the effective user throughput is not only limited by the interference and noise seen at the receiver but also the MAC protocol used to prevent collisions in CSMA/CA networks. Considering the heterogeneous nature of frames transmitted by (or activities of) users on the network, for  a frame size $\digamma \left(\mbox{bits}\right)$, the total transmission time is
\begin{equation}
t_{ij} = \frac{\digamma \left(\mbox{bits}\right)}{ r_{ij} } \left(s\right).
\label{tTime}
\end{equation}


This frame transmission time assumes an idle channel, no collision or retransmissions occurring. However, in reality, retransmissions and channel busy state could prolong the eventual frame transmission time. Figure~\ref{fig:figTimer} depicts a simple MAC procedure that could potentially affect user throughput where a STA senses a busy channel after receiving CTS and waits for a time interval DIFS, backs off and transmit after another round of RTS/CTS exchange and SIFS waiting time. This MAC procedure could potentially degrade aggregate effective throughput. Therefore, the transmission time $t_{ij}$ in (\ref{tTime}) ought to include the MAC-induced delay:
\begin{equation}
\tau_{ij} = t_{\mbox{DIFS} }+ t_{\mbox{SIFS}} + t_{bf} + t_{ack},
\label{delay}
\end{equation}

\noindent $t_{\mbox{DIFS} }$ is the total DIFS time, $t_{\mbox{SIFS}}$ represents the total SIFS time, $t_{bf} = \frac{CW_{max}}{2} \times \mbox{slot-time}$ is the back-off time given a maximum contention window (CW), $CW_{max}$, while $t_{ack}$ is the total transmission time of the block ACK. Therefore,  we define the \textit{effective throughput} of STA$_i$ through AP$_j$ as
\begin{equation}
\beta_{ij} = \left(\frac{1}{t_{ij} + \tau_{ij}}\right) \cdot \log_2 \Pi, \qquad i \in \mathcal{N}, j \in \mathcal{M},
\label{effectiveThruput}
\end{equation}

\noindent where $\Pi$ depends on the \textit{modulation and coding scheme} (MCS) used for in WLAN systems. For instance, $\Pi = 2$ for BPSK. 

\subsection{Proportional-fair Throughput Utility}

To ensure fair end-to-end performance in terms of the throughput defined in (\ref{effectiveThruput}) for each user, we adopt proportional fairness \cite{jmo}, which has been widely used in existing studies (e.g \cite{weili}, \cite{bethanab}). For a network-wide system performance, the throughput utility function is defined as
\begin{equation}
\tilde{\beta}_{ij} =  \Upsilon_\delta \left(\beta_{ij}\right), \qquad i \in \mathcal{N}, j \in \mathcal{M} 
\label{farbw}
\end{equation}
\noindent where $\delta$ is the parameter that determines the degree of fairness on the network. If all users achieve equal throughputs ($100 \%$ fairness), $\delta =  1$ and substituting (\ref{effectiveThruput}) into (\ref{farbw})
\begin{equation}
\Upsilon_\delta \left(\beta_{ij}\right) = \log \beta_{ij},
\label{delta1}
\end{equation}
\noindent and for any discrepancy in fairness, $\delta \neq 1$, 
\begin{equation}
\Upsilon_\delta \left(\beta_{ij}\right) = \frac{ \beta_{ij}^{1 - \delta} }{ 1 - \delta }, \qquad \forall i \in \mathcal{N}, \delta \geq 0.
\label{delta2}
\end{equation}
\noindent where $\beta_{ij}$ is given in (\ref{farbw}). In practice, it is difficult to achieve a $100\%$ fairness for $\delta = 1$ due to the heterogeneous nature of users traffics or payloads. Therefore, subsequently, we consider only the case when $\delta = 0.5$. 

\section{Proposed User-AP Association Framework}{\label{framework}}

This section begins with the formulation of the user-AP association optimization problem in Section~\ref{optmz}. The optimization problem is then transformed into a bipartite graph matching problem in Section~\ref{proform}, which is solved algorithmically in Sections~\ref{opassoc1} and \ref{sec:dynassoc}.

\subsection{User-AP Association Optimization}{\label{optmz}}

Let $\xi_{i, j}$ denote a binary variable indicating that STA $i$ associates with AP $j$ and $\mathcal{X} \triangleq \{\xi_ {i,j} \} | 1 \leq i \leq {N}, 1\leq j \leq M $ is the set of user-AP associations on the entire network. Our primary objective is to find an optimal set $\mathcal{X}^*$ of user-AP associations across the network that maximizes aggregate proportionally fair user throughput. This problem is formulated as the following constrained optimization problem:
\begin{subequations}
	\label{userAssocOptimization}
	\begin{align}
	& \text{maximize} & & \sum_{j = 1}^{M} \sum_{i = 1}^{N} \tilde{\beta}_ {ij} \cdot \xi_ {ij} ,  \label{objectveFunc} \\
	& \text{subject to} & & \sum_{j = 1}^{M} \xi_{ij} = 1  \qquad \forall i \in \mathcal{N} \label{assocConstraint}\\
	& & & \sum_{j = 1}^{M}\xi_{ij}\mbox{SINR}_{ij} \geq \gamma \qquad \forall i \in \mathcal{N}, \label{sinrConsr}\\
	& & & \xi_{ij} \in \{0,1\}, i \in \mathcal{N}, j \in \mathcal{A}
	\label{asso}
	\end{align}
\end{subequations}

\noindent where in (\ref{asso}) $\xi_{ij}  \in \{0,1\} = 1$ if STA$_{i}$ is associated with AP$_{j}$, and zero otherwise. Constraint (\ref{assocConstraint}) ensures that each STA associates with exactly one AP, while (\ref{sinrConsr}) represents the link SINR constraint, respectively. With the assumption that STAs receive sufficiently strong signal from densely deployed APs, constraint (\ref{sinrConsr}) is already taken into account and becomes redundant; this implies that every user obtains minimum SINR required to support the basic WLAN rate. A feasible solution to problem (\ref{userAssocOptimization}) is obtainable through search for a set of user associations that maximizes the aggregate effective throughput.

\begin{rem}\label{downlink-remark}
	The $\mbox{SINR}_{ij}$ satisfying constraint (\ref{sinrConsr}) is the best uplink SINR of STA$_i$ when associated with any of the APs within range and we refer to the AP offering this best SINR as the best serving AP. Using uplink SINR as a metric in distributing users among APs only captures STA interference. To capture AP interference in the downlink, Problem~(\ref{userAssocOptimization}) can be reformulated by modifying constraint (\ref{sinrConsr}) to take downlink SINR into account. {This problem is formulated in \cite{psdb} to take downlink SINR into account}. 
\end{rem}

\subsection{User-AP Association on Graph}{\label{proform}}

The user-AP association problem (Eqn.~(\ref{userAssocOptimization})) in itself is a \textit{combinatorial problem}. We next show that it can be transformed to a simpler maximum weighted bipartite graph matching problem. To cast problem (\ref{userAssocOptimization}) as a graph matching problem that depends only on finding association set $\mathcal{X} \triangleq \{\xi_ {i,j} \}$, the SINR constraints in (\ref{sinrConsr}) are relaxed to:
\begin{equation}
\xi_{ij}\mbox{SINR}_{ij} \geq \gamma, \qquad i \in \mathcal{N}, j \in \mathcal{A},
\end{equation}
\noindent and the optimal solution $\xi^*_{ij}$ is determined for each STA $i \in \mathcal{N}$:
\begin{equation}
j^* = \underset{j}{\arg\max} \{ \tilde{\beta}_{ij}: j \in {{\mathcal{A}}}, \mbox{SINR}_{ij} \geq \gamma  \}, 
\end{equation}
\noindent setting $ \xi_{i{j^*}} = 1 $ and $ \xi_{ij} = 0$ $ \forall j \in \mathcal{A}\setminus\{ {j^*} \}$, the rate upper bound for STA$_i$ is therefore:
\begin{equation}
\mathbf{\beta}^u= \tilde{\beta}_{i{j^*}}.
\label{upp}
\end{equation}

\begin{figure}[!h]
	\centering
	\includegraphics[width=3in]{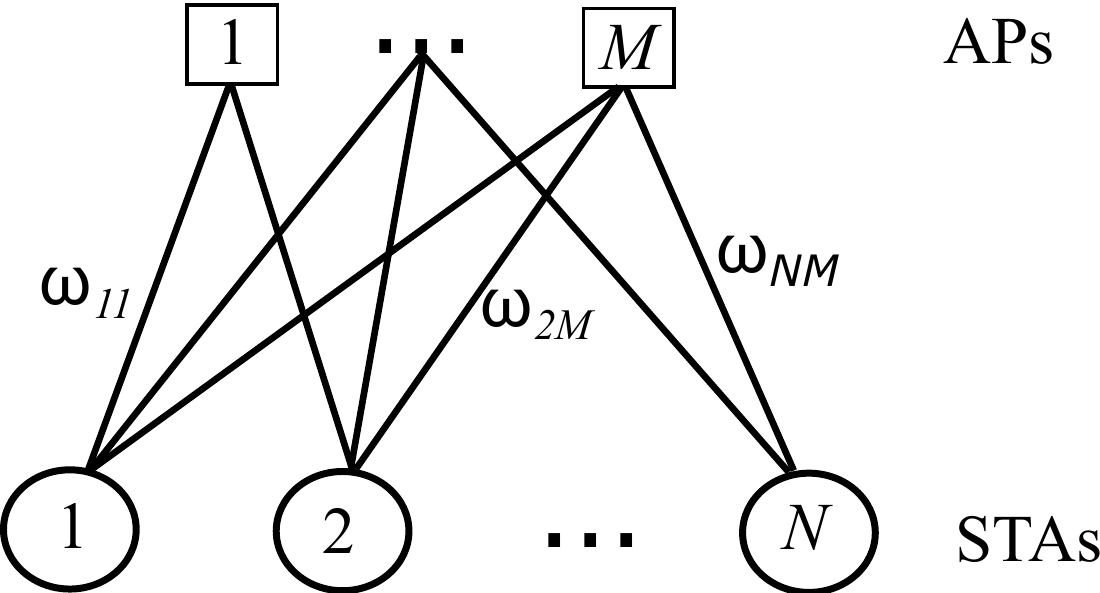}
	\caption{Graph representation of a wireless LAN.}
	\label{fig:fig2}
\end{figure}

Consequently, if we let $\tilde{\beta}_{ij}$ represent the graph edge weight $w_{ij}$, i.e. $w_{ij}  \triangleq \tilde{\beta}_{ij} $, then problem (\ref{userAssocOptimization}) becomes a combinatorial search for ${j^*}$ for each $i \in \mathcal{N}$, which is achieved by using the proposed algorithm in Section~\ref{opassoc1}. Figure~\ref{fig:fig2} represent a WLAN bipartite graph $ G = \left({\mathcal{N}}, {\mathcal{A}}, {\mathcal{E}}\right) $ in which ${\mathcal{N}} = \{1, 2, \cdots, N\}$ is a non-empty set of $N$ STAs at the bottom, ${\mathcal{A}} = \{1, 2, \cdots, M\}$ is the set of $M$ APs, and $ {\mathcal{E}} \subseteq {\mathcal{N}} \times {\mathcal{A}}$ is the set of edge weights. Each STA $i \in {\mathcal{N}}$ can be matched to exactly one AP such that the aggregate effective throughput in (\ref{objectveFunc}) is maximized. A match $ {\mathcal{M}} \subseteq \mathcal{E}$ connects each vertex in ${\mathcal{N}} \cup {\mathcal{A}}$ such that each vertex in $\mathcal{A}$ is an endpoint of at least one edge. 

An optimal association occurs if each STA vertex is connected to exactly one AP vertex in $\mathcal{A}$ that offers $\mathbf{\beta}^u$ for each STA. In the dense WLAN scenario considered, full coverage is assumed, i.e., each AP vertex is connected to at least one STA vertex in ${\mathcal{N}}$. The edge $\left(i, j\right)$ from STA $i$ to AP $j$ is  associated with weight $w_{ij}$. Bipartite graph matching has been applied in different domains \cite{ref10},\cite{ref11},\cite{ref14}. Here, it is used to obtain a match (or association) that maximizes aggregate throughput. The semi-matching approach, which is a relaxation of the maximum bipartite matching problem \cite{ref14}, is employed. A semi-matching is a subset of edges $ {\mathcal{M}}\subseteq {\mathcal{E}}$ such that each vertex in $\mathcal{ A}$ is an endpoint of at least one edge in $\mathcal{M}$. That is, each AP is an endpoint of at least one STA; the objective is to find an improved set of associations, a matching $\mathcal{M}$ that maximizes the aggregate throughput utility.

Next, we introduce a throughput-optimal centralized user-AP association algorithm to solve the graph problem in Section~\ref{proform} for quasi-static and  dynamic networks. The algorithm presented in Section~\ref{opassoc1} for the quasi-static network is offline, while the algorithm presented in Section~\ref{sec:dynassoc} can be implemented dynamically. For the algorithms in Sections~\ref{opassoc1} and \ref{sec:dynassoc}, we assume the proportional fairness parameter $\delta$ is known \textit{a-priori} or chosen arbitrarily. 

\subsection{Graph-based Association Algorithm (GAA)}\label{opassoc1}

Using graph theory, the optimization problem in (\ref{userAssocOptimization}) is equivalent to a maximum weighted bipartite graph matching problem of finding the set of edge weights that maximizes the objective in (\ref{objectveFunc}) or yields (\ref{upp}) for all STAs. The Kuhn-Munkres algorithm (KMA) provides an optimal solution to assignment problems in polynomial time \cite{ref10},\cite{ref11},\cite{ref14}. Applied to optimal AP association, the solution is summarized as Algorithm~\ref{algo:optma1}, which employs the classical KMA. Algorithm \ref{algo:optma1} has bipartite graph, $G$, that is constructed with the $N \times M$ weight matrix $\mathcal{W}$ generated from CSI $\mathbf{H}$ and other MAC layer parameters according to $\tilde{\beta}_{ij}$ for each $i \in \mathcal{N}$. To reduce computational complexity, the size of $G$ can be reduced by obtaining $\mathbf{H}$ of each STAs for only APs within range as opposed to obtaining a global CSI for all STAs. An AP is within range of an STA if the received signal strength (RSS) is above the receiver sensitivity threshold. In such scenario, the algorithm does not need CSI of an STA through all APs' links in the network.

\begin{figure}[!p]
	\removelatexerror
	\begin{algorithm}[H]
		\scriptsize
		\KwIn{$G = \left({\mathcal{N}}, {\mathcal{A}}, {\mathcal{E}}\right)$, $\delta$, $\mathbf{H}$, $\mathbf{V}$, $\mathbf{W}$}
		\KwOut{Match ${\mathcal{M}}$ of $N$-STAs to $M$-APs.} 
		\textbf{1. Initialization:} \;
		\For{$i \leq N $}{ \For{$ j \leq M $}{  \If{  $\mbox{RSS} > \phi $ }{ $\mbox{Compute } \mbox{SINR}_{ij} $ using (\ref{sinrmeasure}) } } }
		\textbf{2. Compute Edge weights:}\\
			\For{$i \leq N $}{ \For{$ j \leq M $}{  \If{  $\mbox{RSS} > \phi $ }{ \If{$\delta = 1 $}{ Compute $\tilde{\beta}_{ij} $ using (\ref{delta1}) \\ $\omega_{ij} \leftarrow \tilde{\beta}_{ij}; \mathcal{W} \leftarrow \omega_{ij} $ } \Else{ Compute $\tilde{\beta}_{ij} $ using (\ref{delta2}) \\ $\omega_{ij} \leftarrow \tilde{\beta}_{ij}; \mathcal{W} \leftarrow \omega_{ij} $ } } \Else{ $\omega_{ij} \leftarrow 0 $ (AP not in range)} } }
		\textbf{3. Association Optimization:} $G = \left({\mathcal{N}}, {\mathcal{A}}, {\mathcal{W}}\right)$\;
		Begin with an empty match, ${\mathcal{M}} \leftarrow \emptyset$ \label{matchn}\;
		Perform steps in $\mathbf{ KMA-Routine}$ with $\mathcal{W}$ as input\;
		Output optimal matching, ${\mathcal{M}}$\\
		\hrulefill \\
		$\mathbf{ KMA-Routine}$\\ Let $\mathcal{C} $ be the number of rows  and columns with covered zeros.\\
		\hrulefill \\
		\textbf{Input}: $N \times M$ weight matrix $ \mathcal{W}$.\\
		\textbf{Step 1.} Set $\gamma_{\mbox{r-min}}$ as the row minimum in $\mathcal{W}$. Subtract $\gamma_{\mbox{r-min}}$ from each row: \\ \hspace{\algorithmicindent} \hspace{\algorithmicindent} \hspace{\algorithmicindent} $\omega_{ij} = \omega_{ij} - \gamma_{\mbox{r-min}}$\\
		\textbf{Step 2.} Set $\gamma_{\mbox{c-min}}$ as the column mini. in $\mathcal{W}$. Subtract $\gamma_{\mbox{c-min}}$ from each column: \\ \hspace{\algorithmicindent} \hspace{\algorithmicindent} \hspace{\algorithmicindent} $\omega_{ji} = w_{ji} - \gamma_{\mbox{c-min}}$\\
		\textbf{Step 3.} Obtain new weight matrix $\mathcal{W}^*$ with changes from \textbf{Steps 1} and \textbf{2}.\\
		\textbf{Step 4.} Cover all zeros in $\mathcal{W}^*$.  \\
		\textbf{Step 5.} Add the minimum of the uncovered elements to covered elements in $\mathcal{W}^*$, \\ \hspace{\algorithmicindent}  \hspace{\algorithmicindent} then subtract the minimum element $\omega_{m}$ from each element: \\ \hspace{\algorithmicindent} \hspace{\algorithmicindent} \hspace{\algorithmicindent} $\omega_{ij} = \omega_{ij} - \omega_{m}$\\
		\textbf{Step 6.} If $\mathcal{C} \neq N$. Repeat \textbf{Step 5}. Else go to \textbf{Step 7}\\
		\textbf{Step 7.} Obtain optimal association (matching), ${\mathcal{M}}$, by selecting a zero in each of \\ \hspace{\algorithmicindent} \hspace{\algorithmicindent} 
		the $i$ rows such that each row has only one selected zero.\\
		\textbf{Step 8.} Choose the AP in the corresponding $j^{th}$ column of $i^{th}$ in \textbf{Step 7}, which \\ \hspace{\algorithmicindent} \hspace{\algorithmicindent} indicates the associated AP of STA $i \in {\mathcal{N}}$.
		\caption{Graph-based Association Algorithm (GAA)}
		\label{algo:optma1}
	\end{algorithm}
\end{figure}

The search for the maximum matching is achieved on line \ref{matchn} of Algorithm~\ref{algo:optma1} by performing the weighted bipartite matching using KMA with $\mathcal{W}$ as the input, which produces an optimal matching $\mathcal{M}$. The optimal user-AP associations in  $\mathcal{M}$ solves (\ref{objectveFunc}) for a given value, $\delta$, for proportional fairness. With centralized execution of Algorithm~\ref{algo:optma1}, optimal AP association is achievable with polynomial complexity $O\left(N^3\right)$ on a WLAN with $N$ STAs. Efficient implementation of the Kuhn-Munkres algorithm has been extensively discussed in the literature, e.g., \cite{lawler}, \cite{steiglitz}, and \cite{mills} and show that slack variables yield efficient $O\left(N\right)$ computation complexity in searching for an augmenting path. The optimal matching set, ${\mathcal{M}}$ is obtained in Algorithm~\ref{algo:optma1} and the selected edge weights solve (\ref{objectveFunc}). For this type of \textit{combinatorial problem}, it is well-known that KMA achieves optimal matching \cite{lawler}, \cite{steiglitz}, and \cite{mills}, hence, Algorithm~\ref{algo:optma1} produces an optimal user-AP association set. 

\subsection{Graph-based Dynamic Association Algorithm (GDA)}\label{sec:dynassoc}

We next discuss a dynamic implementation of  Algorithm~\ref{algo:optma1} for scenarios where STAs are joining and leaving the network, i.e., a method to update the AP association as the network changes. This is not an issue in the IEEE 802.11 standard SSF association scheme since the association process is determined  by simply choosing the AP with the strongest RSS without considering achievable rate or interference at the selected AP. Two types of dynamic situations need to be considered. First, new users may join and others may exit the network. Second, there may be user mobility within the network. The proposed online STA-AP association algorithm is motivated by \cite{mills},\cite{ismail} and we will adopt the notation therein. Given an initial optimized association, ${\mathcal{M}}$ from Section~\ref{opassoc1}, we seek to update the assignment of STAs to APs when edge weights change due to mobility, addition of new users, or exit of users; the procedures to address these dynamics are highlighted in Algorithm~\ref{dynamicsAlgo}.

\subsubsection{New STA Admissions}\label{newstas}

As the network receives new STAs, vertices are added to the existing weighted graph $G$ to create an extended graph $G^{'}$. The incremental assignment algorithm \cite{ismail} solves the maximum-weighted matching problem of the extended graph with $O\left(N^{2}\right)$ complexity to produce a new matching $\mathcal{M}^{'}$. Given the following definition:
\begin{defin}
	Given the original weighted bipartite graph $G$ with optimal matching $\mathcal{M}$, $\mathcal{M}^{'}$ is the optimal matching of an extended graph $G^{'}$ containing the new pair of vertices (STAs) not included in $G$.
\end{defin}

Let $G$ denote the graph of the optimal association obtained in Section~\ref{opassoc1} with optimal matching ${\mathcal{M}}$. When a new vertex is added to $G$, we create an extended graph $G^{'} = \left({\mathcal{N}}^{'}, {\mathcal{A}}^{'}, {\mathcal{E}}^{'}\right)$ with weight matrix ${\mathcal{W}}^{'}$ of size $N+1 \times N+1$ containing new pairs of vertices. The goal is to determine the maximum-weighted matching of the extended graph $G^{'}$  that gives optimal association without affecting the throughputs of existing STAs. As shown in Algorithm~\ref{dynamicsAlgo}, subsequent to the initialization phase on lines 2 and 3, the algorithm executes lines \ref{incstart} - \ref{incend} if there is an extended or new graph $G^{'}$ due to new users entering the network or users exit.

\begin{figure}[!h]
	\removelatexerror
	\begin{algorithm}[H]
		\caption{Graph-based Dynamic Assoc. Alg. (GDA)}
		\label{dynamicsAlgo}
		\scriptsize
		
		\KwIn{$G = \left({\mathcal{N}}, {\mathcal{A}}, {\mathcal{E}}\right)$, optimal matching $\mathcal{M}$ from Alg.~\ref{algo:optma1}, $\delta$, $\mathbf{H}$, $\mathbf{V}$, $\mathbf{W}$}
		\KwOut{Maximum-weighted matching ${\mathcal{M}}^{'}$ or ${\mathcal{M}}^{*}$}
		
		
		Set extended graph $G^{'} = \left({\mathcal{N}}^{'}, {\mathcal{A}}^{'}, {\mathcal{E}}^{'}\right)$, $N^{'} = |{\mathcal{N}}^{'}|$ \label{incgraph}\;
		OR an affected graph $G^{*} = \left({\mathcal{N}}^{*}, {\mathcal{A}}, {\mathcal{E}}\right)$,  ${\mathcal{M}}$, $ \mathcal{W}^* \leftarrow \mathcal{W} $ \label{dyngraph}
		
		\uIf{$N^{'} = N + 1$ $|| N - 1 || N^{'} \neq N$ \label{incstart}}{
			${\mathcal{M}}^{'}$ $\leftarrow $ $ {\mathcal{M}} $ \{begin with previous optimal matching $\mathcal{M}$\} \\
			
			Construct $N+1 \times N+1$ weight matrix ${\mathcal{W}}^{'}$:\\
			\hspace{\algorithmicindent}\hspace{\algorithmicindent}$ w_{ij}^{'} \leftarrow \tilde{\beta}_{ij}$,  $\mathcal{W}^{'}$ $= \{w_{ij}^{'} | \forall i \in {\mathcal{N}}^{'}, j \in {\mathcal{A}}^{'}\}$ \\
			
			Set dual variables $\alpha_{N+1}$ and $\theta_{N+1}$:\\
			$\theta_{N+1} = \max \left\{ \max\limits_{1 \leq i \leq N} \left( w_{i\left(N+1\right)}^{'} - \alpha_{i}\right), w_{\left(N+1\right),\left(N+1\right)}^{'} \right\}$ \label{dualbeta}\;
			
			$\qquad \alpha_{N+1} = \max\limits_{1\leq j \leq N+1} \left(w_{\left(N+1\right) j}^{'} - \theta_{j}\right)$ \label{dualalpha}\\
			
			Execute \textit{OMP}($\mathcal{W}^{'}$, $\mathcal{M}$, $\theta_{N+1}$, $\alpha_{N+1}$) function on line \ref{omp}\;
			Output the new optimal matching $\mathcal{M^{'}}$. \label{incend}
		} 
		\uElseIf{$N^* = N$ \&\& $\mathcal{W}^* \subset \mathcal{W}$ \label{dynstart}}{
			\hspace{\algorithmicindent} ${\mathcal{M}}^{*} \leftarrow {\mathcal{M}}$  \{begin with previous optimal matching $\mathcal{M}$\}\\
			\uIf{ row $j$ of $\mathcal{W}$ changed: }{ 
				Set ${\mathcal{M}}^{*}$ = ${\mathcal{M}}$ - ${\mathcal{E}}$$\left(i, j\right)$ \{Remove the changed or affected edge\}\;
				Set $\alpha_{i}^{*} = \max\limits_{j}\left(w_{ij} -  \theta_{j}\right)$\;
				$\mathcal{W}^{*} \leftarrow w^*_{ij} $\{update the affected edge ${\mathcal{E}}$$\left(i, j\right)$ \}
			}
			\uElseIf{column $i$ of $\mathcal{W}$ changed}{
				Set ${\mathcal{M}}^{*}$ = ${\mathcal{M}}^{*}$ - ${\mathcal{E}}$$\left(i, j\right)$ \textit{// Remove the changed or affected edge}\\
				Set $\theta_{j}^{*} = \max\limits_{i}\left(w_{ij} -  \alpha_{i}\right)$ 
			}\Else{
				Exit \{No changes in edge weights\}
			}
			
			Execute \textit{OMP}($\mathcal{W}^{*}$, $\mathcal{M}$, $\theta_{j}^{*}$, $ \alpha_{i}^{*} $) function on line \ref{omp} \label{ompdyn}\;
			Output the new optimal matching $\mathcal{M}^*$ \label{dynend}\;
		}
		\Else{
			Exit \{No changes in edge weights\}
		}
		
		\Fn{OMP ($\mathcal{W}$, $\mathcal{M}$, $\theta$, $\alpha$) \label{omp}}{Optimal Matching Procedure (OMP):
			
			Begin with matching $\mathcal{M}$ from Alg.~\ref{algo:optma1}
			
			Perform one iteration of the KMA routine \textbf{Stage} in \cite[Fig. 3]{ mills } \\
		}
		
	\end{algorithm}
\end{figure}

Given the new graph $G^{'}$ on line~\ref{incgraph}, a new weight matrix $\mathcal{W}^{'}$ is constructed and two \textit{dual variables} $\alpha$ and $\beta$ are defined. The dual variables are initial labels for each vertex in $G^{'}$ because the dual of the assignment problem (\ref{userAssocOptimization}) is feasible when $\alpha_i + \theta_j \leq w_{ij} \forall i \in \mathcal{N}, j \in \mathcal{A}$. On lines \ref{dualbeta} and \ref{dualalpha}, feasible values are determined for the dual variables $\theta_{N+1}$ and $\alpha_{N+1}$ of the newly added vertices. The previous matching $\mathcal{M}$ from Alg.~\ref{algo:optma1}, the new weight matrix $\mathcal{W}^{'}$, and the dual variables $\theta_{N+1}$ and $\alpha_{N+1}$ are inputs to the optimal matching procedure (OMP) function on line~\ref{omp}, which produce a new optimal matching $\mathcal{M}^{'}$ using the KMA routine \textbf{Stage} in \cite[Fig. 3]{ mills }. The initialization phase is an $O\left(N\right)$ operation and by executing one iteration of the KMA routine \cite[Fig. 3]{ mills }, new matching $\mathcal{M}^{'}$ is obtainable from lines \ref{incstart} - \ref{incend} with complexity $O\left(N^{2}\right)$.

\subsubsection{User Mobility}\label{mobility}

User mobility within the network and arrivals of new users result in a dynamic scenario. When STAs join the network, there is a possibility of increased interference to other users, and consequently, UL SINRs of some links may decrease. In such situations, we update ${\mathcal{M}}$ to obtain a new optimal association ${\mathcal{M}}^{*}$ when any of the link SINRs (edge weights) change either due to user mobility and/or increased link interference. A change in the weight matrix $\mathcal{W}$ resulting from a change in the $i^{th}$ column and/or $j^{th}$ row necessitates the need for a new association set ${\mathcal{M}}^{*}$. This is accomplished using lines~\ref{dynstart} - \ref{dynend} of Algorithm~\ref{dynamicsAlgo} following the initialization of the new graph $G^*$ on line~\ref{dyngraph}. 

In other words, a change in a user's weight may alter the optimal matching $\mathcal{M}$ from Algorithm~\ref{algo:optma1} and thus, we need to efficiently find an optimal solution to the new problem with the SINR weights $\mathcal{W}^*$. Basically, a new feasible value is computed for the dual variable $\alpha_i$ if the weight change occurs in row $j$ or for the dual variable $\alpha_j$ if the change occurs in column $i$. Once the new feasible value is determined, the search for a new optimal matching $\mathcal{M}^*$ is accomplished by executing the OMP function on line~\ref{ompdyn} with the updated weight matrix $\mathcal{W}^*$, previous matching $\mathcal{M}$ and the new dual variables $ \alpha_{i}^{*} $ and $ \alpha_{i}^{*} $ as inputs. If there are $N^{*}$ affected nodes, $N^{*}$ stages of the KMA routine \cite{mills}  are required with overall computational complexity $O\left(N^{*}N^{2}\right)$. In summary, the above dynamic association algorithms are computationally preferable over the offline version but with diminishing gain as the proportion of changed nodes increases.
\begin{rem}
	For the dynamic algorithm, a change could occur in weight matrix $\mathcal{W}$. We could start from $\mathcal{M}$ to get $\mathcal{M}^{'}$ and go from $\mathcal{M}$ to ${\mathcal{M}}^*$. The search for a new optimal match $\mathcal{M}$ could start from the previous optimal matching $\mathcal{M}$. Depending on the network state, we could go from optimal matching $\mathcal{M}^{'}$ to a new optimal matching $\mathcal{M}^*$. It is assumed that the CSI of the new users becomes available as they join the network.
\end{rem}

\section{Performance Evaluation}\label{performanceI}

In this section, simulation results are presented to evaluate the performance of the proposed user-AP association algorithm in a dense \textit{interference-dominated} WLAN environment with uplink MIMO. To simulate a dense WLAN, we assume that the distributions of STAs and APs on the network follows Poisson point processes (PPP) with intensities that represent the densities of STAs and APs. For the uplink transmissions, STAs use the distribution  coordination function (DCF) 802.11 MAC protocol and the PHY is based on the OFDM scheme in 802.11ax. To simulate WLAN, we develop a custom detailed CSMA/CA protocol simulator and perform a Monte Carlo simulation with $10^4$ realizations of the network.

\begin{table*}[!b]
	\centering
	\caption{Simulation parameters}
	\label{table:simpara}
	\begin{tabular}{|l|l|l|l|l|}
		\toprule
		\hline
		\multicolumn{2}{|c|}{\textbf{PHY Parameters}} & \multirow{4}{*}{} & \multicolumn{2}{c|}{\textbf{MAC Parameters}} \\ \cline{1-2} \cline{4-5} 
		Simulation network area 	&    200m $\times$ 200m      
		&                   &      MAC Payload (Packet size)     &    1500 bytes       \\ \cline{1-2} \cline{4-5} 
		Number of APs, $M $	&     $|\mathcal{A}|$      &                   &    SIFS, $t_{\mbox{SIFS}}$     &   $10 \mu s$         \\ \cline{1-2} \cline{4-5} Number of STAs $N  $	&      $|\mathcal{N}|$     &                   &    ACK time, $ t_{ack}$   &   64 $\mu s$     \\ 
		\cline{1-2} \cline{4-5} AP and STA Densities	&      $\eta_m$ and $\eta_n$     &                   &     DIFS, $t_{\mbox{DIFS}}$       &     SIFS + 2 $\times$ Slot-time      \\ 
		\cline{1-2} \cline{4-5} Channel Bandwidth	&      $20 $ $\mbox{MHz}$     &                   &      CCA threshold    &     -70 dBm      \\
		\cline{1-2} \cline{4-5} MCS	&      BPSK     &                   &       Slot-time    &     20 $\mu \mbox{s}$      \\ 
		\cline{1-2} \cline{4-5} OFDM FFT Size	&      $64$     &                   &        Contention Window (CW), $\max$   &      1024    \\ 
		\cline{1-2} \cline{4-5} Guard Interval	&      $800 \mbox{ ns}$     &                   &       Contention Window (CW), $\min$    &     32      \\
		\cline{1-2} \cline{4-5} OFDM Symbol Duration	&     12.8 $\mu s$  + 0.8/1.6/3.2 $\mu s$ CP     &                   &     Packet arrival rate      &    1/Slot-time        \\ 
		\cline{1-2} \cline{4-5} Noise Floor	&      -100 dBm/Hz     &                   &          RTS/CTS &    Enabled       \\
		\cline{1-2} \cline{4-5} STA Transmit power	&      15.85mW ($12 \mbox{dBm}$)     &                   &     Receiver sensitivity, $\phi$       &     -75 dBm      \\
		\cline{1-2} \cline{4-5} Subcarrier Spacing	&      78.125 kHz     &                  &    MAC Header       &      22 bytes   \\\hline
	\end{tabular}
\end{table*}

\begin{figure}[!h]
	\centering
	\includegraphics[width=4in, height=3in]{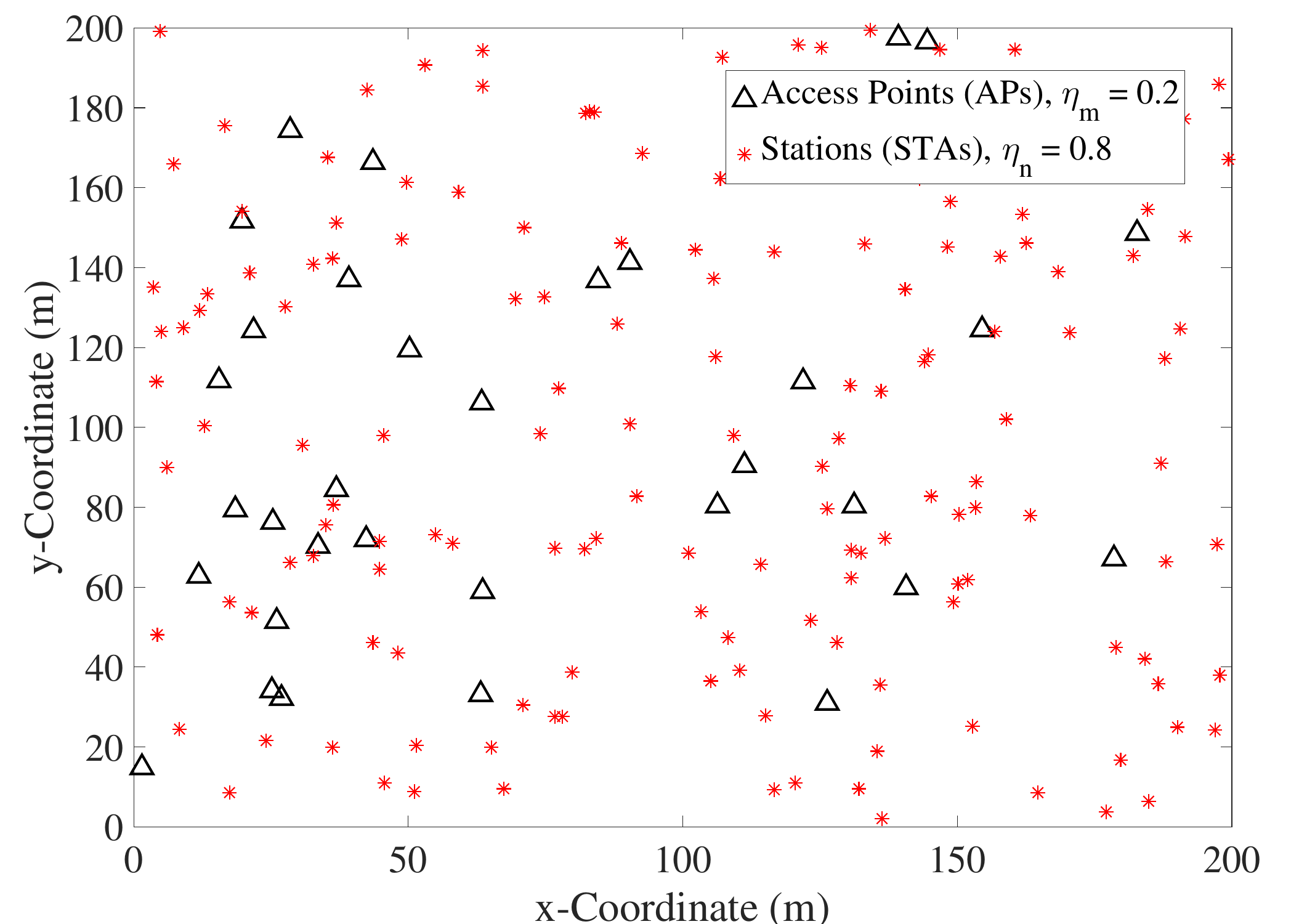}
	\caption{Sample realization of a network with an AP density $\eta_m = 0.2$ and an STA density $\eta_n = 0.8$: $M = |\mathcal{A}|= 35$, $N = |\mathcal{N}|= 169$.}
	\label{netwrk}
\end{figure}


\subsection{Experimental Setup}

To model a high density WLAN, the locations of STAs and APs are modeled as realizations of Poisson point processes (PPPs) due to the uncertainty in the nodes' locations. The simulated WLAN for this experiment consists of $N = |\mathcal{N}|$ STAs and $M = |\mathcal{A}|$ APs where $\mathcal{N}$ and $\mathcal{A}$ represent the PPPs of the STAs and the APs with intensities $\eta_n$ and $\eta_m$, respectively. Figure~\ref{netwrk} depicts a simulated distribution of STAs and APs using PPPs. Unless otherwise stated, each STA is equipped with $U = 4$ transmit antennas while each AP has $K = 8$ receiving antennas. The key PHY and MAC parameters used in this experiment are based on the IEEE 802.11ax standard and are summarized in Table~\ref{table:simpara}. In simulating the PHY, Orthogonal Frequency Division Multiplexing (OFDM) is implemented with FFT size of 64, channel bandwidth 20-MHz, and a guard interval of 800 ns. The transmit power of each STA is $15.85 \mbox{ mW} $ ($12 \mbox{dBm}$) while the noise floor of $-100$ dBm/Hz is used to generate the noise power spectral density. We consider only the 2.4 GHz bands, and the received power is modeled using the log-distance path-loss model with ITU path loss exponent of 3.4 for \textit{rush-hour} propagation \cite{ref27}. 

For channel sensing, the CCA sensitivity (or threshold) $\Gamma$ is set to $-70 \mbox{dBm}$. Interference at an AP is the measure of interference power received from concurrent transmitters (STAs) located outside of the carrier sensing range (CSR) of the desired signal. The CSR depends on the CCA sensitivity threshold and for a CCA threshold of $-70 \mbox{dBm}$, the CSR is approximately $80$m. Setting the CCA threshold and the CSR enables us to capture the set ${\mathcal{N}}_{\mbox{csma}}$ of interference sources in (\ref{sinrmeasure}). This interference model is based on the operation of the CSMA/CA protocol and generally, STAs in the same BSS would adhere to the CSMA/CA back-off rule. 

Assuming that each AP has full knowledge of the CSI through the use of \textit{pilot signals} or channel sounding, each AP constructs and feeds back the beamforming matrix $\mathbf{W}$ to the STAs for precoding. Although, the proposed algorithms are independent of the beamforming technique used. For the MAC and channel contention procedure, the DCF protocol is implemented for access mechanism. The SIFS value for 2.4-GHz  is set to $t_{\mbox{SIFS}} = 10 \mu s$. Each STA generates an identical packet size of 1500 bytes, and the generated packet arrives with exponential distribution at each STA's buffer at the mean rate of 1 packet per slot-time and each slot-time is $20 \mu s$ long. Other MAC parameters such as maximum and minimum contention window (CW) and the DIFS are as stated in Table~\ref{table:simpara}, and each ACK frame takes $ t_{ack} = 64 \mu s$.

\subsection{Simulation Results and Performance Benchmarking}

Using the aggregate \textit{effective throughput} as the performance metric, simulations are performed to quantify the performance of the proposed user-AP association scheme. We benchmark the performance of the proposed algorithm in Section~\ref{framework} with the conventional SSF association scheme used in current WLAN systems, the best performance first (BPF) \cite{weili}, the \textit{greedy algorithm} \cite{yzhangd} and the \textit{SmartAssoc algorithm} \cite{smartasoc}. In the BPF algorithm, an STA selects an AP with best performance, which is determined using a \textit{performance revenue function} \cite[Eqn. 9]{weili}. The BPF algorithm assumes a weight for each user, and herein it is assumed that all users have equal weight. With the \textit{greedy algorithm}, an STA measures the load at the APs and selects an AP with the minimum load. The \textit{SmartAssoc algorithm} simply allows an STA to estimate the load at each candidate AP and selects an AP such that the \textit{norm} of the loads at the candidate APs is minimized. Herein, this candidate list contains those APs that offer an RSSI value of at least -75 dBm (the minimum receiver sensitivity). The load definitions for \textit{Greedy} and \textit{SmartAssoc} algorithms can be found in \cite{yzhangd} and \cite{smartasoc}, respectively, and we adopt these AP load definitions to depend on the channel rate in (\ref{ratePHY}).

\begin{figure}[!h]
	\centering
	\includegraphics[width=5in, height=4in]{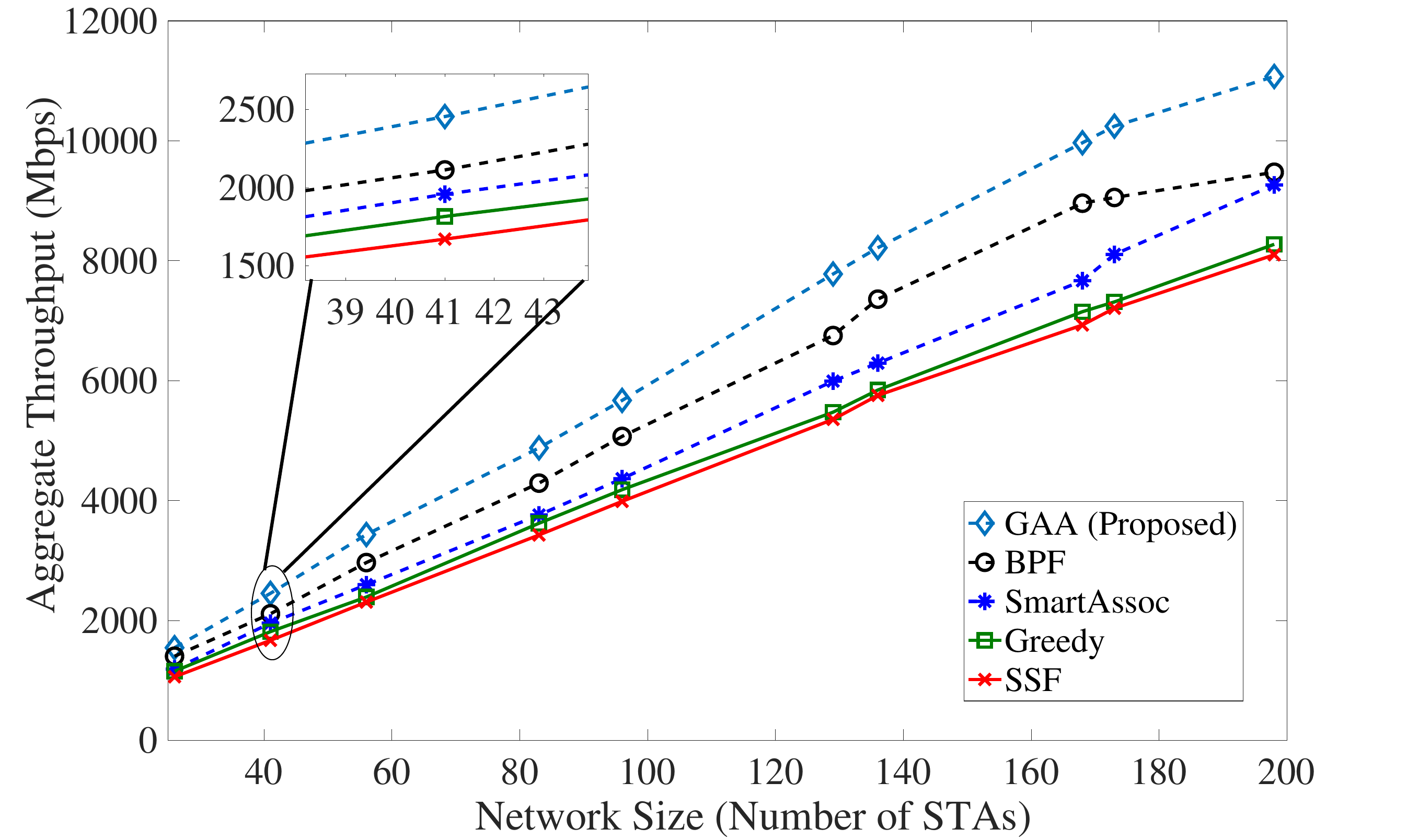}
	\caption{Aggregate throughput versus network size with $K = 8$ antennas at the APs, $U = 4$ transmit antennas at the STAs, fairness index $\delta = 0.5$ and $M = |\mathcal{A}| = 35$.}
	\label{res1}
\end{figure}

Figure~\ref{res1} shows the aggregate throughput with respect to the network size given $\delta = 0.5$ for a monte carlo simulation of $10^4$ slot-times. Figure~\ref{res1} depicts the achievable sum \textit{throughput} when users and APs are equipped with $U = 4$ and $K = 8$ antennas, respectively. Examining Fig.~\ref{res1} at a network size of 100 STAs, the proposed GAA algorithm achieves $46.9 \%$, $35.1 \%$, $25.4 \%$ and $16.1 \%$ gains over the existing SSF, \textit{Greedy}, \textit{SmartAssoc} and BPF algorithms, respectively. With increasing network density up to 198 contending STAs, GAA has $36.7\%$, $33.8\%$, $19.5\%$ and $16.8 \%$ greater throughput than the SSF, \textit{Greedy}, \textit{SmartAssoc} and BPF algorithms, respectively. This significant performance gain is primarily due to the fact that the strongest RSS and the load-based schemes (\textit{Greedy} and \textit{SmartAssoc}) consider APs with best RSS and minimum load, respectively, which are not necessarily offering the best performance in terms of end-to-end throughput. 

   \begin{figure}[!h]
	\centering
	\includegraphics[width=5in, height=4in]{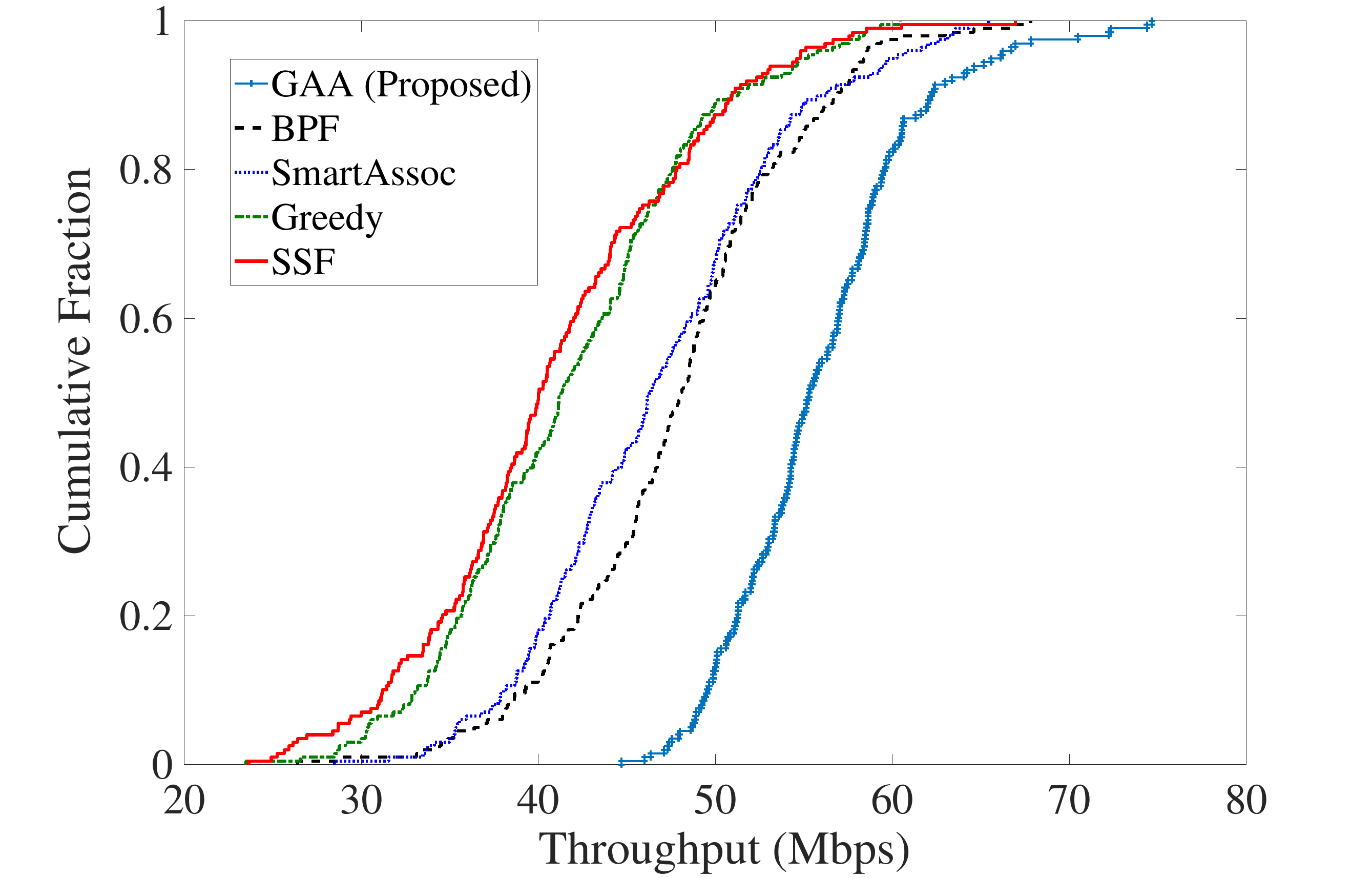}
	\caption{Cumulative distribution of users throughput with $K = 8$, $U = 2$, $\delta = 0.5$, $N = |\mathcal{N}|= 190$ and $M = |\mathcal{A}| = 35$.}
	\label{res2}
\end{figure}

Figure~\ref{res2} captures the distribution of per-user throughput for a network with 198 STAs and 35 APs where STAs and APs have $U = 2$ and $K = 8$ antennas, respectively. For the $10th$ percentile of STAs, the proposed scheme outperforms the existing algorithms. More precisely, at $10th$ percentile of STAs, using the proposed GAA, $22.7 \%$, $28.1 \%$, $49.6 \%$ and $59.1 \%$ throughput improvement is achievable compared to BPF, \textit{SmartAssoc}, \textit{Greedy} and SSF schemes, respectively. Similarly, significant throughput improvement is apparent at the $90th$ percentile. These performance gains at both $10th$ and $90th$ percentiles imply that the proposed framework (GAA) could potentially improve per-user throughput for the users with poor and those with good links.

 \begin{table*}[!t]
	\centering
	\scriptsize
	\caption{Summary of results: Aggregate throughput versus percentage gain}
	\label{resultsumm}
	\begin{tabular}{|c|c|c|c|c|c|c|l|c|c|c|c|}
		\hline
		\multicolumn{1}{|l|}{}                       & \multicolumn{1}{l|}{}                                   & \multicolumn{5}{c|}{{ \textbf{Aggregate Throughput (Mbps)}}}                                                                                         &                    & \multicolumn{4}{c|}{{ \textbf{\% Gain by GAA (Proposed) over}}}                                                                                         \\ \cline{1-7} \cline{9-12} 
		\textbf{STA Density} & \textbf{No. of Contending STAs} & \textbf{SSF} & \textbf{GAA} & \textbf{SmartAssoc} & \textbf{Greedy} & \textbf{BPF} &                    & \textbf{SSF} & \textbf{SmartAssoc} & \textbf{Greedy} & \multicolumn{1}{l|}{\textbf{BPF}} \\ \cline{1-7} \cline{9-12} 
		0.1     & 18   & 638    & 882     & 757    & 690      & 807                                &                    & 38.2                                & 16.5                                          & 27.8                                   & 9.3                                                         \\ \cline{1-7} \cline{9-12} 
		0.2       & 40      & 1498      & 1977       & 1734        & 1514   & 1761                           &      & 32.0             & 14.0                                       & 30.6                                   & 12.3                                                         \\ \cline{1-7} \cline{9-12} 
		0.3                                          & 64                                                      & 2309                                & 3104                                  & 2617                                       & 2309                                      & 2802                                  &                    & 34.4                                & 18.6                                       & 34.4                                   & 10.8                                                         \\ \cline{1-7} \cline{9-12} 
		0.4                                          & 77                                                      & 2808                               & 3846                                & 3088                                       & 2905                                  & 3475                                &                    & 37.0                                & 24.5                                       & 32.4                                   & 10.7                                                         \\ \cline{1-7} \cline{9-12} 
		0.5                                          & 95                                                     & 3488                                & 4739                                & 3876                                       & 3559                                  & 4258                                &                    & 35.9                               & 22.3                                       & 33.2                                   & 11.3                                                        \\ \cline{1-7} \cline{9-12} 
		0.6                                          & 114                                                    & 4001                                & 5598                                 & 4583                                       & 4172                                   & 4990                                 &                    & 39.9                                & 22.1                                       & 34.2                                   & 12.2                                                        \\ \cline{1-7} \cline{9-12} 
		0.7                                          & 142                                                    & 5187                                & 7188                                 & 5837                                       & 5270                                  & 6502                                 &                    & 38.6                                & 23.1                                       & 36.4                                   & 10.6                                                        \\ \cline{1-7} \cline{9-12} 
		0.8                                          & 169                                                    & 6069                                & 8395                                 & 6967                                       & 6312                                   &    7447                             &                    & 38.3                                & 20.5                                       & 33.0                                   & 12.7                                                        \\ \cline{1-7} \cline{9-12} 
		0.9                                          & 187                                                    & 6677                                & 9188                                 & 7647                                      & 6754                                   & 8274                                 &                    & 37.6                               & 20.2                                       & 36.0                                   & 11.0                                                        \\ \cline{1-7} \cline{9-12} 
		1                                            & 194                                                     & 7083                                & 9704                                 & 7950                                       & 7084                                   & 8668                                & \multirow{-9}{*}{} & 37.0                                & 22.1                                       & 37.0                                   & 12.0                                                         \\ \hline
	\end{tabular}
\end{table*}
 
\begin{figure}
	\centering
	\includegraphics[width=5in, height=4in]{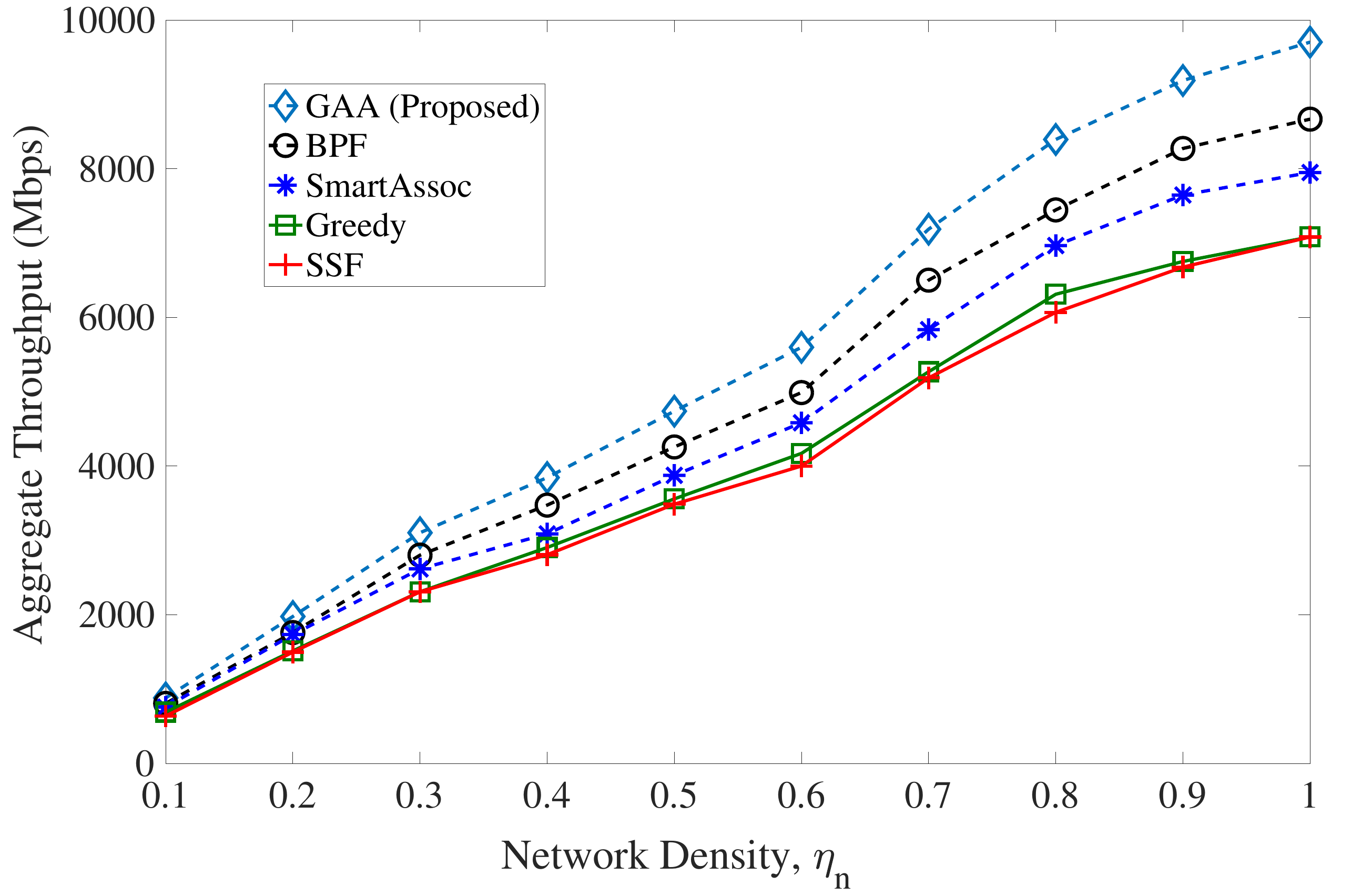}
	\caption{Aggregate throughput versus network density with $K = 8$, $U = 2$, $\delta = 0.5$.}
	\label{res3}
\end{figure}

Table~\ref{resultsumm} summarizes the simulation results in this study. Given various densities and the corresponding numbers of STAs, the percentage gain of GAA over each of the existing algorithms is presented. From all simulated network density scenarios, the results in Table~\ref{resultsumm} show that on average, aggregate network throughput improvement of GAA is  $36.9 \%$, $20.4 \%$, $33.5 \%$ and $11.3 \%$ compared to SSF, \textit{Greedy}, \textit{SmartAssoc} and BPF. In Figure~\ref{res3}, while increasing the network density $\eta_n$, the aggregate network throughput of GAA still shows improvement over the existing schemes. From all tested network densities, it is apparent that the throughput improvement of GAA remains plausible even as the network density grows. 


\subsection{Simulated Dynamic Scenario}

In many cases, wireless networks are not \textit{quasi-static}, since new users join or existing users leave the network sporadically. In addition, user mobility creates dynamics in the network. When new users join the network, the interference distribution changes and so does aggregate throughput. In the case when a user leaves the network, the interference level is expected to reduce or at least the remaining users could transmit at an improved rate. Hence, in performance assessment of dynamic networks, we focus on the impact of new entrants on the aggregate throughput. Although, GDA has identical performance to GAA, it is a dynamic implementation of GAA and possesses lower complexity in terms of updating user-AP association when network changes.

\begin{figure}[!ht]
	\centering
	\includegraphics[width=5in, height=4in]{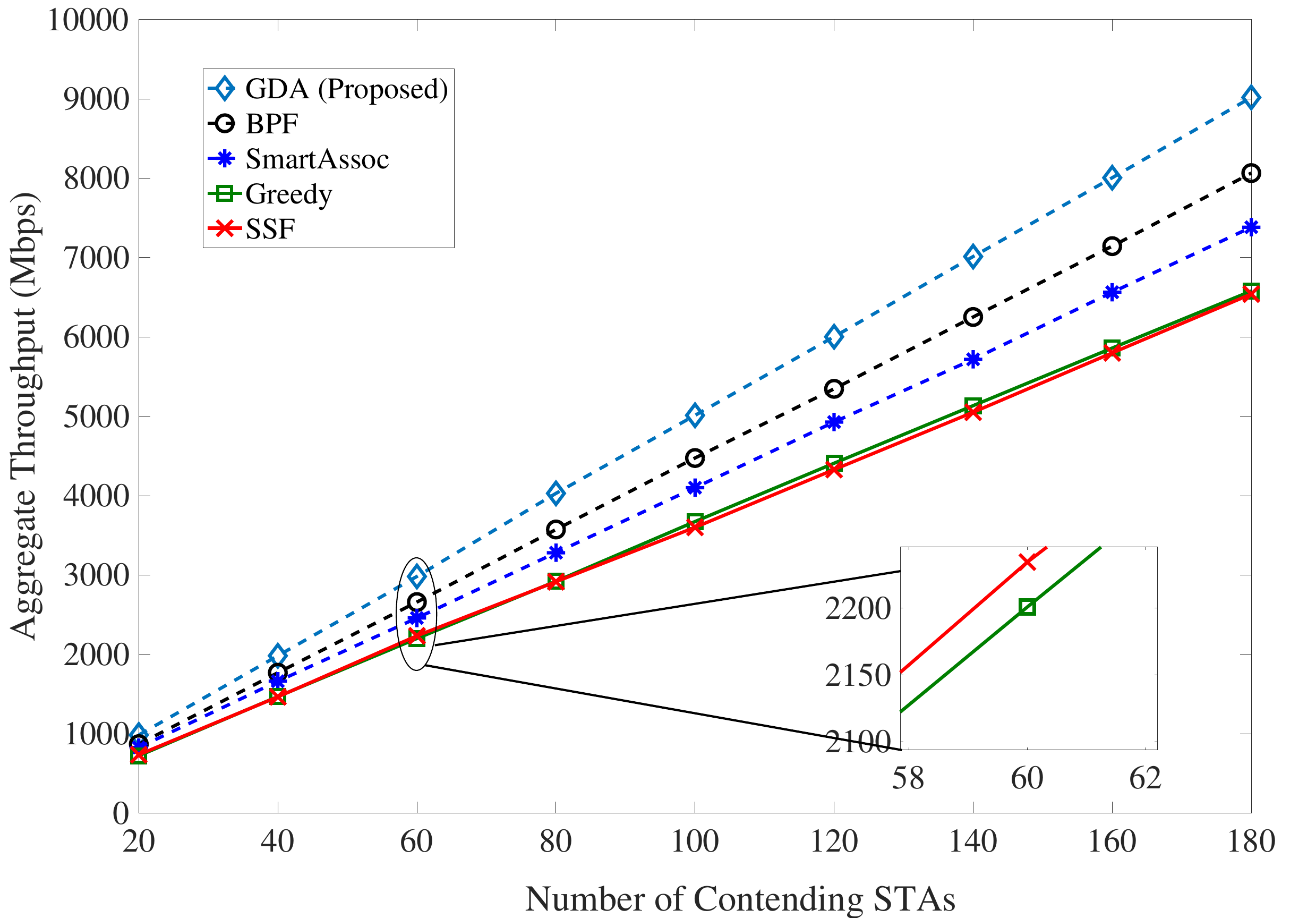}
	\caption{Aggregated effective throughput of GDA vs. network size with $K = 8$, $U = 2$ and $\delta = 0.5$. }
	\label{resdyn1}
\end{figure}

In Figure~\ref{resdyn1}, starting the network with 20 STAs, the number of users is increased randomly at each time-slot and a fraction of the new entrants change location within the network according to the \textit{Random Waypoint Mobility Model} (RWMM) \cite{waypoint}. Figure~\ref{resdyn1} depicts aggregate throughput as density of users on the network increases due to new entrants. Examining the aggregate throughput when network size reaches 80 STAs, the throughput improvements of GDA are $12.7\%$, $ 22.8 \%$, $ 37.9 \%$  and $38.2 \%$ over the BPF, the SmartAssoc, the Greedy, and the SSF algorithms, respectively. For network size of 180 STAs, significant gain over existing algorithms becomes apparent. The main reason for low throughput gain when there are fewer STAs  (for instance, 20 STAs in Figure~\ref{resdyn1}) is that interference and channel correlation are negligible when there are fewer contending STAs. Hence, selecting AP based on some metrics other than strongest signal has less impact on the throughput. However, as the network density grows, it is paramount to take other factors (interference, channel correlation, MAC protocol effects etc.) into account when selecting an AP.

\section{Conclusions}\label{conclusion}

Major challenges in high density WLAN include interference management, user-AP coordination and persistent channel contention/collision. An algorithm is proposed that coordinates user-AP associations considering CSI knowledge, effects of MAC protocol and interference at the target APs. A dynamic implementation of the proposed algorithm is also introduced to more efficiently handle arrivals of new users and user mobility. Simulation is performed to benchmark the performance of the proposed scheme to existing algorithms. Our primary observation in this study is that coordinating user-AP associations using channel knowledge and MAC protocol effect, could significantly improve the aggregate network throughput. From all simulated  interference-dominated UL-MIMO scenarios, the proposed scheme achieves an average of $36.9 \%$, $33.5 \%$, $20.4 \%$ and $11.3 \%$  gains over SSF, \textit{Greedy}, \textit{SmartAssoc} and BPF, respectively. Although, as the user density grows, computational complexity of the proposed algorithm increases as $O\left(N^3\right)$ for $N$ STAs, an improvement in the aggregate throughput is obtainable and is a potentially attractive tradeoff for more throughput in dense networks.

\end{document}